\documentclass[twocolumn,eqsecnum,showpacs,preprintnumbers,amssymb,aps]{revtex4}
\usepackage{graphicx}
\usepackage{dcolumn}

\begin{document}

\title{Theoretical studies of the atomic transitions in boron-like ions: \\
Mg VIII, Si X and S XII}
\vspace{0.5cm}

\author{$^a$H. S. Nataraj \protect \footnote[2] {Electronic address: nataraj@iiap.res.in}, $^b$B. K. Sahoo, $^a$B. P. Das, $^a$R. K. Chaudhuri and $^c$D. Mukherjee \linebreak 
$^a${\it Non-Accelerator Particle Physics Group, Indian Institute of Astrophysics, Bangalore-34, India}\linebreak 
$^b${\it Max Planck Institute for the Physics of Complex Systems, N\"othnitzer Str. 38, D-01187 Dresden, Germany} \linebreak
$^c${\it Department of Physical Chemistry, Indian Association for Cultivation of Science, Kolkata-700 032, India}}
 \noaffiliation
\begin{abstract}
\noindent
In this paper, we have carried out the calculations of the weighted oscillator strengths 
and the transition probabilities for a few low-lying transitions of boron-like ions: Mg VIII, 
Si X and S XII which are astrophysically important, particularly, in the 
atmospheres of the solar corona.
We have employed an all-order relativistic many-body theory called the relativistic 
coupled-cluster theory to calculate very precisely these atomic quantities of astrophysical interest.
We have reported for the first time the transition probabilities for some forbidden transitions which are unavailable in the literature; either theoretically or experimentally. We also discuss the physical effects associated with these transitions. Our data can be used for the identification of spectral lines arising from the coronal atmospheres of Sun and Sun-like stars having an extended corona.
\end{abstract}
\date{Received date; Accepted date}
\vskip1.0cm

\keywords{atomic data - methods: analytical - Sun: X-rays}
\pacs{31.15.Ar, 31.15.Dv, 32.30.Jc, 31.25.Jf, 32.10.Fn}

\maketitle
\section{Introduction}
With the remarkable advances in the field of observational astronomy like the
deployment of satellite probes for data acquisition, there is a considerable
interest for accurate calculations of the oscillator strengths and the transition
probabilities for highly stripped ions which are very important in astrophysics,
mainly in the identification of spectral lines \cite{fawcett75, fawcett87, podobedova,reconditi,murcray}.
Various electro-magnetic transitions from the low-lying single valence 
excited states, $2s^2\,2p_{3/2}\,(^2P_{3/2})$, $2s^2\,3s\,(^2S_{1/2})$, 
$2s^2\,3d_{3/2}\, (^2D_{3/2})$, and $2s^2\,3d_{5/2}\, (^2D_{5/2})$ to the 
ground state in the highly ionized boron-like ions such as $Mg^{7+}$, $Si^{9+}$ and 
$S^{11+}$ are observed in the solar atmosphere \cite{jordan69,elraju75}. Most of the lines
correspond to the soft X-ray waveband and have the potential to probe the
chromosphere-corona transition region and possibly the coronal hole regions of 
the solar atmosphere \cite{jordan69,flower75a,flower75b}, where the 
temperatures would be of the order of a million Kelvin. The relative line 
intensity
ratios in Mg VIII and Si X line emission spectrum have been found to be density
sensitive \cite{dwivedi80,vernazza}. The EUV line intensity ratios of these ions have 
been studied \cite{bhatia} to infer the electron density in different solar 
features such as active region, quiet sun and off-limb. Therefore, lines emitted
from boron-like ions can be used as a powerful tool in the diagnostics of the 
electron density \cite{vrmason78} and the temperature in the solar atmosphere.
Interestingly, the soft X-ray coronal emission lines of S XII in 
the stellar binary Capella, which is one of the nearby Sun-like stars, have been observed by Audard et al. \cite{audard} using the 
high resolution RGS XMM-Newton satellite.  \\ 

There are a few calculations 
of certain transition probabilities of the considered boron-like ions available in the literature; some of them are completely non-relativistic and
based on the multi-configuration Hartree-Fock (MCHF) method \cite{shamey71,garstang62,dankwort76};
some others are based on MCHF calculations with Breit-Pauli corrections (MCHF+BP) \cite{dankwort78,fischer83,tachiev} and there are a few calculations based on the relativistic many-body perturbation theory (MBPT) \cite{safronova} and relativistic multi-reference configuration interaction (MRCI) method \cite{konard}. Often the theoretical calculations are scaled to match the 
observed transition energies \cite{elraju75,tachiev}. Given the increasing need for the accurate spectroscopic data in astrophysics, it is necessary to use the  all-order relativistic many-body
 methods like the relativistic coupled-cluster (RCC) theory \cite{lindgren} to calculate the principal atomic quantities of astrophysical interest such as the energy levels, the oscillator strengths, the transition 
probabilities and the lifetimes of the excited states.  

The study of boron-like ions is interesting from the point of view of the
strong core-valence electron correlation effects and also because the transition energies are in the reach of current laboratory astrophysics experimental facilities like electron beam ion trap (eBIT) \cite{lapierre}. In this paper, we present 
both the allowed and forbidden transition amplitudes and the corresponding
transition probabilities of a few low-lying states in the boron-like ions. We also discuss the behaviour of 
correlation effects associated with these calculations.

The organization of the paper is as follows: In section II, we give the working formulas
for the transition probabilities and the oscillator strengths and briefly discuss the RCC method employed in
calculating these quantities. In section III, we 
present the results and compare them with those available in the literature and the conclusions are drawn in the last section.
\section {Theory and method of calculation}
The spontaneous transition probabilities due to E1, E2, M1, and M2 operators from a state $\vert J_f M_f\rangle$ to the state $\vert J_i M_i\rangle$ are given by \cite{shore},
\begin{eqnarray}
A^{E1}_{J_fJ_i} = \frac{64 \pi^4 e^2a_0^2}{3h\lambda^3(2J_f + 1)} S^{E1} = \frac{2.0261\times 10^{-6}}{\lambda^3(2J_f + 1)} S^{E1} ,
\end{eqnarray}
\begin{eqnarray}
A^{E2}_{J_fJ_i} = \frac{64 \pi^6 e^2a_0^4}{15h\lambda^5(2J_f + 1)} S^{E2} = \frac{1.12\times 10^{-22}}{\lambda^5(2J_f + 1)} S^{E2} ,
\end{eqnarray}
\begin{eqnarray}
A^{M1}_{J_fJ_i} = \frac{64 \pi^4 e^2a_0^2 (\alpha/2)^2}{3h\lambda^3(2J_f + 1)} S^{M1} = \frac{2.6971\times 10^{-11}}{\lambda^3(2J_f + 1)} S^{M1} ,
\end{eqnarray}
and
\begin{eqnarray}
A^{M2}_{J_fJ_i} = \frac{64 \pi^6 e^2a_0^4 (\alpha/2)^2}{15h\lambda^5(2J_f + 1)} S^{M2} = \frac{1.491\times 10^{-27}}{\lambda^5(2J_f + 1)} S^{M2} ,
\end{eqnarray}
respectively, where, the numerical factor applies for the wavelength $\lambda$ 
in {\emph cm} and the transition
line strength $S^{O}$, defined as the absolute square of the transition matrix 
element i.e.
\begin{eqnarray}
 S^{\,O}_{(J_f;J_i)}  = \;  \mid {\langle J_f \,\vert \vert\, O \,\vert \vert\, J_i \rangle} \mid^2
\end{eqnarray}
where, $\langle J_f \,\vert \vert\, O\, \vert \vert\, J_i \rangle$ is the reduced
matrix element for the appropriate multipole operator $O$,
in atomic units (au). Here, $J$ is the total angular momentum quantum number.

The single particle
reduced matrix elements for the E1 and E2 operators in length gauge and the gauge independent M1 and M2 operators are, respectively, given by \cite{johnson1995},
\begin{widetext}
\begin{eqnarray}
\langle \kappa_f\, ||\, e1\, ||\, \kappa_i \rangle
&=& \frac{3}{k}\, \langle \kappa_f \,||\, C_q^{(1)} \,||\, \kappa_i \rangle
\int_0^{\infty} \left\{ j_1(kr)\left[P_f(r) P_i(r) + Q_f(r) Q_i(r)\right] \right. \nonumber \\
&+&j_2(kr) \left. \left[\frac{\kappa_f - \kappa_i}{2}\, [P_f(r) Q_i(r) + Q_f(r) P_i(r)] 
+ [P_f(r) Q_i(r) - Q_f(r) P_i(r)]\right] \right\} dr,
\label{e1red}
\end{eqnarray}
\begin{eqnarray}
\langle \kappa_f\, ||\,e2\,||\,\kappa_i \rangle &=& \frac {15}{k^2} \, \langle \kappa_f\, ||\,C^{(2)}\,||\, \kappa_i \rangle \int_0^{\infty} \left\{ j_2(kr)[P_f(r)P_i(r)+Q_f(r)Q_i(r)] \right.\nonumber \\
&+&j_3(kr)\left. \left[\frac {\kappa_f-\kappa_i}{3}\,[P_f(r)Q_i(r)+Q_f(r)P_i(r)]
+ [P_f(r)Q_i(r)-Q_f(r)P_i(r)]\right] \right\} dr,
\label{e2red}
\end{eqnarray}
\begin{eqnarray}
\langle  \kappa_f\, ||\,m1\,||\, \kappa_i \rangle &=& \frac{6}{\alpha k}\, \langle - \kappa_f\, ||\,C^{(1)}\,||\,\kappa_i \rangle \int_0^{\infty} \frac {\kappa_f+\kappa_i}{2}\, j_1(kr)\, \left[P_f(r)Q_i(r)+Q_f(r)P_i(r)\right] dr,
\label{m1red}
\end{eqnarray}
and
\begin{eqnarray}
\langle \kappa_f\, ||\,m2\,||\,\kappa_i \rangle &=& \frac {30}{\alpha k^2}\, \langle -\kappa_f\, ||\,C^{(2)}\,||\, \kappa_i \rangle \int_0^{\infty} \frac {\kappa_f+\kappa_i}{3}\, j_2(kr)\, \left[P_f(r)Q_i(r)+Q_f(r)P_i(r)\right] dr,
\label{m2red}
\end{eqnarray}
\end{widetext}
where, $\kappa$ is the relativistic angular momentum
quantum number. The radial functions, $P_i(r)$ and $Q_i(r)$ are
the large and small components of the $i^{th}$ single
particle Dirac orbital, respectively. The coefficients of the Racah tensor are given by,
\begin{widetext}
\begin{eqnarray}
\langle \kappa_f \, ||\, C^{(\gamma)}\,||\, \kappa_i \rangle &=& (-1)^{j_f+1/2} \sqrt{(2j_f+1)(2j_i+1)}
                          \left ( \matrix {
                              j_f & \gamma & j_i \cr
                              1/2 & 0 & -1/2 \cr
                                       }
                            \right ) \pi(l_f,\gamma,l_i),
\label{eqn8}
\end{eqnarray}
where $j$ is the single particle angular momentum quantum number. The parity selection rule is given by,
\begin{equation}
  \pi(l1,l2,l3) =
  \left\{\begin{array}{ll}
      \displaystyle
      1 & \mbox{for } l1+l2+l3= \mbox{even}
         \\ [2ex]
      \displaystyle
        0 & \mbox{otherwise.}
    \end{array}\right.
\end{equation}
\end{widetext}
In equations (\ref{e1red}) through (\ref{m2red}), we define the wave vector $k$ as, $k=w\alpha$, where $w=\epsilon_i-\epsilon_j$ is the excitation
energy at the single particle level, $\alpha$ is the fine structure constant, $l$
is the orbital angular momentum quantum number and
$j_n(kr)$ is a spherical Bessel function of order $n$. Since $kr$ is sufficiently
small, we apply the following approximation to calculate the above matrix
elements:
\begin{eqnarray}
j_n(z) \approx \frac {z^n}{1.3.5...(2n+1)}.
\label{eqn9}
\end{eqnarray}
The oscillator strength and the corresponding transition probability for a transition of any multipole type are related by the general formula,
\begin{eqnarray}\label{fvalue}
f_{(J_f;J_i)} = 1.4992\times 10^{-16} A_{(J_f;J_i)} \frac{g_f}{g_i}\lambda^2
\end{eqnarray}
where, $g_f$ and $g_i$ are the degeneracies associated with the final and initial states respectively, $\lambda$ is the wavelength in {\emph \AA} and $A_{(J_f;J_i)}$ is the transition probability in $s^{-1}$.

Generally, in the astrophysical context, one uses the weighted oscillator
strength which is the product of the degeneracy of the initial state and the
oscillator strength and is symmetric with respect to the initial and final 
states; i.e.
\begin{eqnarray}
 gf = (2J_i+1)f_{if} = - (2J_f+1)f_{fi} .
\end{eqnarray}
As it can be inferred from the above equations, there is a great need for precise values of the transition wavelengths
and the transition line strengths for the accurate determination of the oscillator strengths and the transition probabilities.
This demands for a highly powerful many-body method which would include fully relativistic effects and the electron
correlation effects that are sufficiently large especially in many-electron systems. So we employ the RCC theory, which is briefly discussed below, to calculate these quantities of interest.

The starting point of our method is the relativistic generalization of the 
valence universal coupled-cluster (CC) theory introduced by Mukherjee et al 
\cite{mukh,debasish} which was put later in a more compact form by Lindgren 
\cite{lindgren,lind}. In this approach, first we obtain the closed-shell 
Dirac-Fock (DF) wave function ($\vert \Phi_0\rangle$) which corresponds to the electronic configuration $1s^2 \, 2s^2$. This amounts to solving the DF equations for $N-1$ electrons where $N$ is the total number of electrons in the system. These equations can be expressed as,
\begin{eqnarray}
\left[c \vec \alpha_i \cdot \vec p_i + (\beta_i - 1)c^2 + V_{Nuc}(r_i) + U_{DF}(r_i)\right] \vert \phi_i \rangle = \epsilon_i \vert \phi_i \rangle
\label{eqn1}
\end{eqnarray}
where, $c$ is the speed of light in vacuum, $\vec \alpha$ and $\beta$ are the Dirac 
matrices, $V_{Nuc}(r_i)$ is the nuclear potential and $U_{DF}(r_i)$ is the effective average potential called the Dirac-Fock potential, which is given by,
\begin{eqnarray}
U_{DF} (r_i)\vert \phi_i \rangle = \sum_{j=1}^{N-1} \left[\langle \phi_j \vert \frac{1}{r_{12}} \vert \phi_j \rangle \vert \phi_i \rangle -\langle \phi_j \vert \frac{1}{r_{12}} \vert \phi_i \rangle \vert \phi_j \rangle \right]
\end{eqnarray}
The large and small radial components of the single particle relativistic wave functions are expanded in terms of the Gaussian type orbitals (GTOs) of the form \cite{mohanty89, rajat98},
\begin{eqnarray}
g^L_{\kappa_i}(r) = N^L r^{n_{\kappa_i}} e^{-\zeta_i r^2} 
\label{largecomp}
\end{eqnarray}
and 
\begin{eqnarray}
g^S_{\kappa_i}(r) = N^S \left[\frac{d}{dr} + \frac{k}{r}\right] g^L_{\kappa_i}(r)
\label{smallcomp}
\end{eqnarray}
where, $N^L$ and $N^S$ are the normalization factors for the large and small radial components, respectively, of the one-electron orbitals, $n_{\kappa_i}$ varies for each relativistic symmetry and takes an integer value as 1 for $s_{1/2}$, 2 for $p_{1/2}$ and so on.

In equation (\ref{largecomp}) we have used the even tempering condition for the exponents, i.e,
\begin{eqnarray}
\zeta_i = \zeta_0 \, \eta^{i-1} ~~\text{where}~~ i = 1, 2, 3, ..., n
\end{eqnarray}
where $\zeta_0$ and $\eta$ are the user-defined parameters and $n$ is the size of the basis set. In equation (\ref{smallcomp}), the kinetic balance condition is imposed on the small radial components to avoid the variational collapse of the wave functions into the negative energy continuum \cite{mohanty89}.\\
The differential equations (\ref{eqn1}) become matrix eigenvalue equations of the form \cite{szabo},
\begin{eqnarray}
F\,C = S\,C \, \epsilon
\end{eqnarray}
where F, S, C and $\epsilon$ are the Fock matrix, overlap matrix, eigenvector and the eigenvalue matrix, respectively. This is then transformed to a true eigenvalue problem and it is diagonalized to get the energies (eigenvalues) and the mixing coefficients (eigenvectors) for both the occupied and the virtual orbitals. The virtual orbitals (including the $2p_{1/2}$ valence orbital) obtained by this procedure are clearly generated in the $V^{N-1}$ potential of the frozen core orbitals. The details of this method can be found else where \cite{szabo,roothan}.

The exact wave function 
($\vert \Psi_0 \rangle$) for the corresponding closed-shell system is 
calculated in the RCC theory using,
\begin{eqnarray}
|\Psi_0 \rangle = e^T \vert \Phi_0\rangle
\label{eqn2}
\end{eqnarray}
where, T is the excitation operator for the core orbitals. It  is the sum of all
 single, double, triple, and higher order excitations of occupied electrons. The open-shell reference state for the desired valence electron, $v$ can be written as, $\vert\, \Phi_v\rangle = a_v^{\dagger}\, \vert\,\Phi_0\rangle$; where, $a_v^{\dagger}$ is the creation operator for the valence electron and $\vert\,\Phi_0\rangle$ is a closed-shell reference state which is the Slater determinant representing the ${1s^2\,2s^2}$ configuration where as, $\vert\, \Phi_v\rangle$ is the Slater determinant representing, for example, the ${1s^2\,2s^2\,2p_{1/2}}$ configuration.  

The exact wave function for the open-shell atomic system can be expressed
in the RCC theory as,
\begin{eqnarray}
\vert \,\Psi_v\rangle = e^T\{e^{S_v}\} \,\vert\,\Phi_v\rangle
\label{eqn3}
\end{eqnarray}
where, $S_v$ corresponds to the excitation operator for the valence and valence-core orbitals.
Since the systems considered in the present case contain single valence electron
 in their electronic configurations, the non-linear terms in the expansion of
exponential function of $S_v$ will
not exist and the above wave function ultimately reduces to the form,
\begin{eqnarray}
\vert \,\Psi_v\rangle = e^T\{1+S_v\} \,\vert\,\Phi_v\rangle.
\label{eqn4}
\end{eqnarray}
where, $\{ \}$ indicates that the operator is normal ordered.

Even in the few electron systems, it is not possible to consider all correlated
excitations due to huge requirement of the computer memory. In fact, it has
been found that the CC theory with both single and double excitations 
(CCSD) is quite successful in incorporating the maximum correlation effects. However,
we have considered the CCSD method along with the important triple excitations (CCSD(T) method). The electron
affinity energy ($\Delta E_v$) for the valence electron $v$ and the RCC
operator amplitudes are calculated self-consistently using the following coupled
equations,
\begin{widetext}
\begin{eqnarray}
\langle \Phi_L | \{\widehat{He^T}\} |\Phi_0 \rangle &=& \Delta E_0 \ \delta_{L,0}  \label{eqn6} \\
\langle \Phi_K|\{\widehat{He^T}\}S_v|\Phi_v\rangle &=& - \langle \Phi_K|\{\widehat{He^T}\}|\Phi_v\rangle + \left[\langle \Phi_K|S_v|\Phi_v\rangle + \delta_{K,v}\right]\ \Delta E_v
\label{eqn13}
\end{eqnarray}
\end{widetext}
where, ($ | \Phi_L \rangle$) with $L\,(=1,2)$ represents the singly or doubly excited state from the closed-shell reference (DF) wave
function ($L=0$) and $\Delta E_0$ is the correlation energy for the 
closed-shell system. And, ($ | \Phi_K \rangle$) with $K\,(=1,2)$ denotes the singly or doubly excited state from the single valence reference state ($K=v$). The excitation energies ($EE$) between different states are calculated from the electron affinity energies. 

We calculate the transition matrix elements of any physical operator $O$ by using,
\begin{widetext}
\begin{eqnarray}\label{genop}
\langle\, O\, \rangle_{i \rightarrow f} &=& \frac {\langle \Psi_f \,\vert\, O \,\vert\, \Psi_i \rangle} {\sqrt{\langle\Psi_f\,|\,\Psi_f\rangle\,\langle\Psi_i\,|\,\Psi_i\rangle}} \nonumber \\
&=& \frac {\langle \Phi_f\, |\,\{1+S_f^{\dagger}\}\, \overline O \,\{1 +S_i\}\, |\, \Phi_i\rangle} {\sqrt{\langle \Phi_f\, |\, \{1+S_f^{\dagger}\} e^{T^{\dagger}} e^T \{1+S_f\}|\Phi_f\rangle \langle\Phi_i\, |\,\{1+S_i^{\dagger}\} e^{T^{\dagger}} e^T \{1+S_i\} |\Phi_i\rangle}}
\end{eqnarray}
\end{widetext}
where, $\overline O = e^{T^{\dagger}} O e^T$. First, we compute the operator $\overline O$
as the effective one-body and two-body operators using the generalized Wick's theorem
\cite{lindgren} and later sandwich this between the necessary $S_v$ operators.
It has to be noticed that the fully contracted $\overline O$ does not 
contribute in the present calculations. The contribution from the normalization
of the wave functions ($Norm$) is given by,
\begin{eqnarray}
Norm = \langle \Psi_f\, |\, O\, |\, \Psi_i \rangle \{ \frac {1}{\sqrt{\mathcal{N}_f\,\mathcal{N}_i}}-1 \}
\end{eqnarray}
where, $\mathcal{N}_v = \langle \Phi_v \,|\, e^{T^{\dagger}} e^T \,|\, \Phi_v \rangle\,+\, \langle \Phi_v \,|\, S_v^{\dagger} e^{T^{\dagger}} e^T S_v \,|\, \Phi_v \rangle$ for the valence electron $v\, (= i, f)$.
\section{Results and discussion}
In Table \ref{energy}, we present our calculated $EE$s for the $2s^{2}\,2p_{3/2}$, $2s^{2}\,3s_{1/2}$,  $2s^{2}\,3p_{1/2}$, $2s^{2}\,3p_{3/2}$, $2s^{2}\,3d_{3/2}$ and $2s^{2}\,3d_{5/2}$ states from the ground state $2s^{2}\,2p_{1/2}$ for 
all the considered systems and compared our results with the available National Institute of
Standards and Technology (NIST) database \cite{nist}. Our results, in general, agree very 
well with the measured NIST energies except for the fine structure level of the ground states; i.e. $2s^{2}\,2p_{3/2}$ states. This shows that the higher order relativistic
effects are important for these states and also the quantum electrodynamics (QED)
corrections may be required to match the observed results. In all
the systems considered, $EE$s were not known for the $2s^{2}\,3p_{3/2}$ and 
$2s^{2}\,3p_{1/2}$ states and here we have presented them for the first time, that can be 
used in the astrophysical observations for the identification of spectral lines.

\begin{table}[h]
\caption{Comparison of the calculated excitation energies with the tabulated NIST data. The ground state is $1s^22s^22p_{1/2}$. \label{energy}}
\begin{ruledtabular}
\begin{center}
\begin{tabular}{c c c c}
Atomic  & Upper & \multicolumn{2}{c}{$EE$ ($cm^{-1}$)}\\ 
\cline{3-4}
system         &  states  & This work   &    NIST \cite{nist} \\
\hline
 & & & \\
Mg VIII  & $2s^{2}\,3d_{5/2}$  & 1\,339\,798 & 1\,336\,030 \\
         & $2s^{2}\,3d_{3/2}$  & 1\,339\,584 & 1\,335\,860   \\
         & $2s^{2}\,3p_{3/2}$  & 1\,276\,642 &       \\
         & $2s^{2}\,3p_{1/2}$  & 1\,275\,753 &       \\
         & $2s^{2}\,3s_{1/2}$  & 1\,210\,608 & 1\,210\,690   \\
         & $2s^{2}\,2p_{3/2}$  & 3\,433 & 3\,302     \\
 & & & \\
\hline
 & & & \\
Si X     & $2s^{2}\,3d_{5/2}$  & 1\,980\,585 & 1\,979\,730  \\
         & $2s^{2}\,3d_{3/2}$  & 1\,980\,086 & 1\,979\,260   \\
         & $2s^{2}\,3p_{3/2}$  & 1\,904\,870 &       \\
         & $2s^{2}\,3p_{1/2}$  & 1\,902\,937 &       \\
         & $2s^{2}\,3s_{1/2}$  & 1\,821\,314 & 1\,822\,000  \\
         & $2s^{2}\,2p_{3/2}$  & 7\,261 & 6\,990   \\
 & & & \\
\hline
 & & & \\
S XII    & $2s^{2}\,3d_{5/2}$  & 2\,753\,686  & 2\,748\,100  \\
         & $2s^{2}\,3d_{3/2}$  & 2\,752\,714  & 2\,747\,400  \\
         & $2s^{2}\,3p_{3/2}$  & 2\,659\,966  &     \\
         & $2s^{2}\,3p_{1/2}$  & 2\,656\,319  &      \\
         & $2s^{2}\,3s_{1/2}$  & 2\,559\,004  &       \\
         & $2s^{2}\,2p_{3/2}$  & 13\,465  & 13\,135 \\
 & & & \\
\end{tabular}
\end{center}
\end{ruledtabular}
\end{table}

\begin{table}[h]
\begin{ruledtabular}
\begin{center}
\caption{Weighted oscillator strengths and transition probabilities for the allowed E1 transitions to the ground state. \label{probabilities}}
\centering
\begin{tabular}{c c c c c c}
{Atomic} & {Upper} &\multicolumn{2}{c}{Weighted oscillator } &\multicolumn{2}{c}{Transition}\\
{System} & {State }&\multicolumn{2}{c}{strength (a.u.)} &\multicolumn{2}{c}{probability $(10^{11}s^{-1})$}\\ \cline{3-4} \cline{5-6}
       &           & This work  &  Others &  This work &  Others  \\
\hline
  &   &   &  &  &  \\
Mg VIII & $2s^{2}\,3s_{1/2}$  & 0.051 & 0.052 \cite{tachiev} & 0.251 &   0.255 \cite{tachiev} \\
       & $2s^{2}\,3d_{3/2}$  & 1.188 & 1.205 \cite{tachiev} & 3.555 & 3.589 \cite{tachiev} \\
  &   &   &  &  &  \\
\hline
  &   &   &  &  &  \\
 Si X  & $2s^{2}\,3s_{1/2}$  & 0.048 & 0.049 \cite{tachiev} & 0.529 & 0.541 \cite{tachiev} \\
       &                     &       & 0.044 \cite{dankwort78}&       & 0.481 \cite{dankwort78}\\
       &                     &       & 0.051 \cite{shamey71}  &       & 0.585 \cite{shamey71}\\
       & $2s^{2}\,3d_{3/2}$  & 1.234 & 1.247 \cite{tachiev} & 8.066 & 8.149 \cite{tachiev} \\
       &                     &       & 1.232 \cite{dankwort78} &       & 8.010 \cite{dankwort78}\\
       &                     &       & 1.234 \cite{shamey71}  &       & 8.310 \cite{shamey71}\\
  &   &   &  &  &  \\
\hline
  &   &   &  &  &  \\
 S XII & $2s^{2}\,3s_{1/2}$  & 0.045 &  0.047 \cite{shamey71} &  0.987 & 1.06 \cite{shamey71}  \\
       & $2s^{2}\,3d_{3/2}$  & 1.260 &  1.256 \cite{shamey71} &  15.932& 16.2 \cite{shamey71} \\
       &   &   &  &  &  \\
\end{tabular}
\end{center}
\end{ruledtabular}
\end{table}
We have used the length gauge in the calculation of the transition properties of E1 and E2. In Table \ref{probabilities}, we have reported the weighted oscillator strengths and
the transition probabilities of a few allowed electric dipole transitions 
obtained from the RCC calculations. We observe that the transition probabilities for $2s^{2}\,3d_{3/2} \rightarrow 2s^{2}\,2p_{1/2}$ transitions are larger than that of $2s^{2}\,3s \rightarrow 2s^{2}\,2p_{1/2}$ transitions that may be because, the overlap of radial wave functions
of the former two states is larger than that of the latter two states. We compare our results with the
 MCHF+BP results of \cite{tachiev} in which the 
calculated energies are scaled to match the observed transition energies and with a few non-relativistic results available in the literature \cite{dankwort78,shamey71}.  
Their methods are less powerful than the all-order RCC theory in incorporating the
electron correlation effects and rigorous relativistic effects and also the RCC method has distinct advantages over the former two \cite{lindgren,sahoo,majumder04}. Here, we have used the unscaled computed wavelengths in calculating the transition probabilities.
 
\begin{table}[h]
\begin{ruledtabular}
\begin{center}
\caption{Contribution from the individual terms for E1 transition amplitude (in a.u.)
of $2s^2\,3s \rightarrow 2s^2\,2p_{1/2}$ and $2s^2\,3d_{3/2} \rightarrow 2s^2\,2p_{1/2}$ transitions. \label{transamps}}
\begin{tabular}{l c c c}
 RCC  & \multicolumn {3}{c}{ $2s^2\,3s \rightarrow 2s^2\,2p_{1/2}$ transition amplitudes} \\
\cline{2-4} terms & Mg VIII& Si X  & S XII  \\
\hline
&   &  &  \\
$\overline O $    &  ~ 1.28\,E-1 &~ 9.84\,E-2  &~ 7.93\,E-2  \\
$\overline O\, S_i$& ~-6.52\,E-3 &~ -4.09\,E-3 &~ -3.04\,E-3 \\
$S_f^\dagger\, \overline O$&  ~-5.38\,E-3 &~ -2.68\,E-3 &~ -1.10\,E-3 \\
$S_f^\dagger\, \overline O \,S_i$ &~ 1.90\,E-3 &~ 1.27\,E-3  &~ 9.12\,E-4  \\
$Norm$  &~ 6.14\,E-5 &~ 1.04\,E-4  &~ 1.03\,E-4  \\
&   &  &  \\
\hline
&   &  &  \\
Total  &~ 1.18\,E-1 &~ 9.30\,E-2  &~ 7.62\,E-2  \\
&   &  &  \\
DF &~ 1.28\,E-1 &~ 9.89\,E-2  &~ 7.99\,E-2  \\
\hline
&   &  &  \\
  & \multicolumn {3}{c}{$2s^2\,3d_{3/2} \rightarrow 2s^2\,2p_{1/2}$ transition amplitudes}\\
\hline
&   &  &  \\
$\overline O $     &~ -5.21\,E-1 &~ -4.36\,E-1 &~ -3.75\,E-1\\
$\overline O\, S_i$&~ 1.73\,E-2  &~ 1.20\,E-2  &~ 9.28\,E-3\\
$S_f^\dagger\, \overline O$&~ -3.03\,E-2 &~ -2.36\,E-2 &~ -1.87\,E-2 \\
$S_f^\dagger\, \overline O \,S_i$&~ -3.66\,E-3 &~ -3.06\,E-3 & ~-1.51\,E-3\\
Norm   &~ -2.54\,E-3 &~ -2.34\,E-3 &~-2.09\,E-3\\
&   &  &  \\
\hline
&   &  &  \\
Total  &~ -5.40\,E-1 &~ -4.53\,E-1 &~ -3.88\,E-1\\
&   &  &  \\
DF  &~  -5.32\,E-1 &~ -4.45\,E-1 &~ -3.83\,E-1\\
\end{tabular}
\end{center}
\end{ruledtabular}
\end{table}
In order to understand the correlation effects, we have given explicitly the individual contributions to the electric dipole
transition amplitudes, in Table \ref{transamps}, for the $2s^{2}\,3s \rightarrow 2s^{2}\,2p_{1/2}$ and
$2s^{2}\,3d_{3/2} \rightarrow 2s^{2}\,2p_{1/2}$ transitions, from the following four terms;
$ \langle \Phi_f \,\vert\, \overline O\, \vert\, \Phi_i \rangle$,
$\langle \Phi_f \,\vert\, \overline O\, S_i\,\vert\, \Phi_i \rangle$,
$\langle \Phi_f \,\vert\, S_f^{\dagger}\, \overline O \,\vert\, \Phi_i \rangle$,
$\langle \Phi_f \,\vert\, S_f^{\dagger}\,\overline O\, S_i\,\vert\, \Phi_i \rangle$
which are obtained on expanding equation (\ref{genop}) and the contributions from 
the normalization factor. As expected, the contribution from the term
$ \langle \Phi_f\, \vert\, \overline O\, \vert\, \Phi_i \rangle$ is large compared to the other three terms as it contains the DF term and a few core correlated terms, where as,
$\langle \Phi_f\, \vert\, S_f^{\dagger}\,\overline O\, S_i\,\vert\, \Phi_i \rangle$ is
smaller than the rest as it contains two orders in
the $S_v$ amplitude. The contribution from $\langle \Phi_f\, \vert\, S_f^{\dagger}\, \overline O\, \vert\, \Phi_i \rangle$ is larger compared to $\langle \Phi_f \,\vert\, \overline O\, S_i\,\vert\, \Phi_i \rangle$ in the $2s^{2}\,3s$ state, where as, it is the other way
round in the case of $2s^{2}\,3d_{3/2}$ state. However, the trends for all the three ions are almost the same for any given transition. We have also presented the DF results in the bottom line of the table
in order to emphasize the correlation contributions to the total results.
 The correlation effects for the electric dipole transition amplitudes are
small compared to the DF values and they are negative, thereby reducing
the contribution of DF values in both the cases.

In Table \ref{corr}, we have given the DF
and CCSD(T) results of the important forbidden transition amplitudes due to
M1, E2 and M2 transitions, which are interesting in the astrophysical context. 
We have presented the percentage difference between these results ($\Delta$)
which represent the contribution due to electron correlation effects.
In a recent calculation of M1 transition probabilities in 
B-like ions using MRCI method with QED corrections \cite{tupitsyn}, it was shown that the 
contribution of the inter-electronic interaction correlation is small for 
the transition from $2s^2\,2p_{3/2}$ to the ground state for S XII. However,
we observe that the contribution from the electron correlation effects are
non-negligible in many of the considered transitions. Interestingly, the M1 transition amplitude for the transition from $2s^2\,3p_{3/2} \rightarrow 2s^2\,2p_{1/2}$ has
very large correlation effects which even change the sign of the CCSD(T) result from the DF result.

\begin{table*}[h]
\caption{Computed transition amplitudes and transition probabilities for a selected forbidden transitions.\label{corr}}
\begin{ruledtabular}
\begin{center}
\begin{tabular}{c c c c c c c c c}
Atomic & Upper & Multipole & \multicolumn{3}{c}{Transition amplitude (in au)}& \multicolumn{3}{c}{Transition probability in ($s^{-1}$)}\\ \cline{4-6}\cline{7-9}
{System} &{State}&         &  DF & CCSD(T) &  $\Delta (in \%)$ & Present & \cite{tachiev} & \cite{konard} \\
\hline
& & & & & & & & \\
Mg VIII  & $2s^{2}\,3d_{5/2}$  & M2   & 1.34\,E0  & 1.33\,E0  & -1.13& 1.90\,E+3  &  &\\
         & $2s^{2}\,3d_{3/2}$  & M2  & -4.12\,E-1& -4.01\,E-1& -2.76 & 2.58\,E+2  &   &\\
         & $2s^{2}\,3p_{3/2}$  & E2  & -2.54\,E-1& -2.67\,E-1&  4.98& 6.78\,E+6  &   &\\
         ~& $2s^{2}\,3p_{3/2}$  & M1  & 1.24\,E-3& -2.86\,E-3 & 143.21& 1.15\,E+2  &  &\\
         ~& $2s^{2}\,3p_{1/2}$  & M1  & 2.62\,E-4 & 5.76\,E-4 & 54.44 & 9.28\,E+0  &   &\\
         & $2s^{2}\,2p_{3/2}$  & E2    & -2.96\,E-1& -2.70\,E-1& -9.96 & 9.85\,E-7  &   8.8804\,E-7& 9.61\,E-7\\
         & $2s^{2}\,2p_{3/2}$  & M1    & -1.15\,E0 & -1.12\,E0 & -2.7 & 3.48\,E-1  &   3.2905\,E-1& 3.21\,E-1\\
& & & & & & & & \\
\hline
& & & & & & & & \\
Si X     & $2s^{2}\,3d_{5/2}$  & M2     & 1.13\,E0  & 1.13\,E0  & 0.33 & 9.65\,E+3   &  &\\
         & $2s^{2}\,3d_{3/2}$  & M2    & -3.38\,E-1& -3.34\,E-1& -2.77 & 1.27\,E+3   &  &\\
         & $2s^{2}\,3p_{3/2}$  & E2    & -1.71\,E-1& -1.79\,E-1& 4.89& 2.27\,E+7   &  &\\
         & $2s^{2}\,3p_{3/2}$  & M1     & 1.74\,E-3& -1.49\,E-3 & 217.08& 1.06\,E+2   &  &\\
         & $2s^{2}\,3p_{1/2}$  & M1    & 3.80\,E-4 & 5.91\,E-4 & 57.37 & 7.24\,E+1   &  &\\
         & $2s^{2}\,2p_{3/2}$  & E2    & -2.01\,E-1& -1.84\,E-1& -9.15 & 1.92\,E-5   &  1.7837\,E-5& 1.91\,E-5\\
         & $2s^{2}\,2p_{3/2}$  & M1    & -1.15\,E0 & -1.12\,E0 & -2.54 & 3.27\,E+0   &  3.1475\,E+0& 3.06\,E+0\\
& & & & & & & & \\
\hline
& & & & & & & & \\
S XII    & $2s^{2}\,3d_{5/2}$  & M2     & 9.67\,E-1 & 9.79\,E-1 & 1.14 & 3.77\,E+4 &   &\\
         & $2s^{2}\,3d_{3/2}$  & M2    & -2.96\,E-1& -2.86\,E-1& -3.68 & 4.80\,E+3 &   &\\
         & $2s^{2}\,3p_{3/2}$  & E2    & -1.24\,E-1& -1.30\,E-1& 4.78& 6.28\,E+7 &   &\\
         & $2s^{2}\,3p_{3/2}$  & M1     & 2.33\,E-3& -3.29\,E-4 & 808.1& 1.37\,E+1 &   &\\
         & $2s^{2}\,3p_{1/2}$  & M1    & 5.24\,E-4 & 6.58\,E-4 & 20.4 & 1.10\,E+2 &   &\\
         & $2s^{2}\,2p_{3/2}$  & E2    & -1.46\,E-1& -1.34\,E-1& -8.63 & 2.23\,E-4 &   & 2.29\,E-4\\
         & $2s^{2}\,2p_{3/2}$  & M1    & -1.15\,E0 & -1.13\,E0 & -2.38 & 2.09\,E+1 &   & 2.03\,E+1\\
& & & & & & & & \\
\end{tabular}
\end{center}
\end{ruledtabular}
\end{table*}
We have also presented, in Table \ref{corr}, the transition probabilities calculated using the transition amplitudes and wavelengths obtained using the CCSD(T) method. These
results are compared with other calculated results available in the literature \cite{tachiev,konard}. As seen from the Table \ref{corr}, our results are in good agreement with the coulomb-guage results of \cite{konard}. 
We have presented the results for a few other transitions which have not been studied earlier. One can obtain the
oscillator strengths using the general formula given in equation (\ref{fvalue})
for these transitions using the above results. 

Our results on the transition probabilities and 
lifetimes of the low-lying transitions in the boron iso-electronic ions may be helpful in the near future for the identification of the spectral 
lines in the regions of extremely 
low density plasma such as those present in the coronal atmospheres of Sun and
a few Sun-like stars. They also serve as bench mark results for the laboratory astrophysics experiments, using eBIT.
\section{Conclusion}
We have calculated the weighted oscillator strengths for a few electric dipole 
transitions and the transition probabilities for some low-lying excited states 
of boron-like ions: Mg VIII, Si X, and S XII which are abundant in the solar 
atmosphere using the
relativistic coupled-cluster theory. It is shown that, the contributions of 
electron correlation effects to the transition amplitudes are non-negligible 
in some transitions. Our results in general are in good agreement with
the calculated values available in the
literature; thereby demonstrating the power of this theory to generate
accurate and reliable atomic data for astrophysics.


\begin{thebibliography}{99}
\bibitem{fawcett75}
B. C. Fawcett, At. Data and Nucl. Data Tables {\bf 16}, 135 (1975)
\bibitem{fawcett87}
B. C. Fawcett and R. W. Hayes, Phys. Scr. {\bf 36}, 80 (1987)
\bibitem{podobedova}
L. I. Podobedova, D. E. Kelleher, J. Reader and W. L. Wiese, J. Phys. Chem. Ref. Data {\bf 33}, 495 (2004)
\bibitem{reconditi}
M. Reconditi and E. Oliva, Astron. Astrophys. {\bf 662}, 274 (1993) 
\bibitem{murcray}
F. J. Murcray, A. Goldman, F. H. Murcray, C. M. Bradford, D. G. Murcray, M. T. Coffey and W. G. Mankin, Astrophys. J. {\bf 247}, L97 (1981)
\bibitem{elraju75}
G. Elwert and P. K. Raju, Astrophys. Solar Science {\bf 38}, 369 (1975)
\bibitem{jordan69}
C. Jordan, MNRAS, {\bf 142}, 501 (1969)
\bibitem{flower75a}
D. R. Flower and H. Nussbaumer, Astron. Astrophys. {\bf 45}, 145 (1975)
\bibitem{flower75b}
D. R. Flower and H. Nussbaumer, Astron. Astrophys. {\bf 45}, 349 (1975)
\bibitem{vernazza}
J. E. Vernazza and J. C. Raymond, Astrophys. J. {\bf 228}, L89 (1979)
\bibitem{dwivedi80}
B. N. Dwivedi and P. K. Raju, Solar Physics {\bf 68}, 111 (1980)
\bibitem{bhatia}
A. K. Bhatia  and R. J. Thomas, 1998, Astrophys. J. {\bf 497}, 483 (1998)
\bibitem{vrmason78}
J. E. Vernazza and H. E. Mason, Astrophys. J. {\bf 226}, 720 (1978)
\bibitem{audard}
M. Audard, E. Behar, M. Gudel,  A. J. J. Raassen, D. Porquet, R. Mewe, C. R. Foley and G. E. Bromage, Astron. Astrophys. {\bf 365}, L329 (2001)
\bibitem{shamey71}
L. J. Shamey, J. Opt. Soc. Am. {\bf 61}, 942 (1971)
\bibitem{garstang62}
R. H. Garstang, Ann. Astroph. {\bf 25}, 109 (1962)
\bibitem{dankwort76}
W. Dankwort and E. Trefftz, Astron. Astrophys. {\bf 47}, 365 (1976)
\bibitem{dankwort78}
W. Dankwort and E. Trefftz, Astron. Astrophys. {\bf 65}, 93 (1978)
\bibitem{fischer83}
C. F. Fischer, 1983, J. Phys. B {\bf 16}, 157 (1983)
\bibitem{tachiev}
G. Tachiev and C. F. Fischer, J. Phys. B {\bf 33}, 2419 (2000)
\bibitem{safronova}
U. I. Safronova, W. R. Johnson and A. E. Livingston, Phys. Rev. A {\bf 60}, 996 (1999)
\bibitem{konard} 
K. Koc, J. Phys. B {\bf 36}, L93 (2003)
\bibitem{lindgren}
I. Lindgren and J. Morrison, {\it Atomic Many-Body Theory}, ($2^nd$ edition, Berlin: Springer-Verlag) (1986)
\bibitem{lapierre}
A. Lapierre et al., Phys. Rev. Lett. {\bf 95}, 183001 (2005)
\bibitem{shore}
B. W. Shore and D. H. Menzel, {\it Principles of atomic spectra}, (New York: John Wiley and Sons) (1968)
\bibitem{johnson1995}
W. R. Johnson, D. R. Plante and J. Sapirstein, Advances in At. Mol. and Optical Physics, {\bf 35}, 255
 (1995)
\bibitem{mukh}
D. Mukherjee, R. Moitra and A. Mukhopadhyay, Mol. Physics {\bf 33}, 955 (1977)
\bibitem{debasish}
D. Mukherjee and S. Pal, Adv. Quant. Chem. {\bf 20}, 281 (1989)
\bibitem{lind}
I. Lindgren, {\it A coupled-cluster approach to the many-body pertubation theory for open-shell systems}, In Per-Olov Löwdin and Yngve Öhrn , editors, Atomic, molecular, and solid-state theory, collision phenomena, and computational methods, International Journal of Quantum Chemistry, Quantum Chemistry Symposium {\bf 12}, pages 33-58, John Wiley \& Sons, March 1978.
\bibitem{mohanty89}
Ajay K Mohanty and E. Clementi, Chem. Phys. Lett. {\bf 157}, 348 (1989)
\bibitem{rajat98}
Rajat K. Chaudhuri, Prafulla K. Panda and B. P. Das, Phys. Rev. A {\bf 59}, 1187 (1999)
\bibitem{szabo}
Attila Szabo and Neil S. Ostlund, {\it Modern quantum chemsitry},(New York: Dover Publications) (1996)
\bibitem{roothan}
Y. K. Kim, Phys. Rev. {\bf 154}, 17 (1967); I. P. Grant, J. Phys. B {\bf 19}, 3187 (1986); H. M. Quiney, I. P. Grant and S. Wilson, {\it ibid.} {\bf 20}, 1413 (1987); I. P. Grant and H.M. Quiney, Adv. At. Mol. Phys. {\bf 23}, 37 (1998)
\bibitem{nist} 
http://physics.nist.gov/PhysRefData/ASD/index.html
\bibitem{sahoo} 
B. K. Sahoo, S. Majumder, H. Merlitz, R. K. Chaudhuri, B. P. Das and D. Mukherjee, J. Phys. B {\bf 39}, 355 (2006)
\bibitem{majumder04}
S. Majumder, G. Gopakumar, R. K. Chaudhuri, B. P. Das, H. Merlitz, U. S. Mahapatra and D. Mukherjee, Euro. Phys. J. D {\bf 28}, 3 (2004)
\bibitem{tupitsyn}
I. I. Tupitsyn, A. V. Volotka, D. A. Glazov, V. M. Shabaev, G. Plunien, J. R.  Crespo Lopez Urrutia, A. Lapierre and J. Ullrich, Phys. Rev. A {\bf 72}, 062503 (2005)

\end{thebibliography}
\end{document}